\documentclass{osa-article}

\journal{osajournal}
\articletype{Research Article}
\newcommand{\ket}[1]{\ensuremath{\vert{#1\rangle}}}

\begin{document}

\title{Detecting correlated errors in twin-field quantum key distribution}

\author{B. Panchumarthi, A. Stephens and M. Beck\authormark{*}}

\address{Department of Physics, Reed College, 3203 SE Woodstock Blvd., Portland, OR 97202}
% \author{Author One,\authormark{1} Author Two,\authormark{2,*} and Author Three\authormark{2,3}}

% \address{\authormark{1}Peer Review, Publications Department, Optica Publishing Group, 2010 Massachusetts Avenue NW, Washington, DC 20036, USA\\
% \authormark{2}Publications Department, Optica Publishing Group, 2010 Massachusetts Avenue NW, Washington, DC 20036, USA\\
% \authormark{3}Currently with the Department of Electronic Journals, Optica Publishing Group, 2010 Massachusetts Avenue NW, Washington, DC 20036, USA}

\email{\authormark{*}beckm@reed.edu} %% email address is required



\begin{abstract*}
We experimentally demonstrate that we can detect correlated errors in a twin-field quantum key distribution (TFQKD) system by using a technique that is related to self-consistent tomography. We implement a TFQKD system based on a fiber-Sagnac loop, in which Alice and Bob encode information in the phase of weak coherent states that propagate in opposite directions around the loop. These states interfere as they exit the loop and are detected by a third party, Charlie, who reports the results of their measurements to Alice and Bob. We find that it is possible for Alice and Bob to detect correlated state-preparation and measurement errors while trusting only their own individual states, and without trusting Charlie's measurements. 
\end{abstract*}

%%%%%%%%%%%%%%%%%%%%%%%%%%  Introduction  %%%%%%%%%%%%%%%%%%%%%%%%%%
\section{Introduction}\label{sec:intro}

Quantum-state tomography (QST) estimates the density operator of an unknown quantum state by performing a series of measurements with perfectly calibrated detectors
\cite{smithey_1993b,leonhardt_1997, paris_2004c}.
Quantum-detector tomography (QDT) estimates the positive-operator-valued measure
(POVM) that describes a detector by probing it with a series of perfectly
characterized quantum states \cite{luis_1999,fiurasek_2001,lundeen_2009}. However, it is now appreciated that in real experiments there is no such thing as perfectly calibrated detectors or perfectly characterized states. There are invariably errors in both of these, and we refer to them as state-preparation and measurement (SPAM) errors. As such, techniques have been developed for estimating the quantum operators that characterize both the state and the measurements, in a self-consistent manner \cite{mogilevtsev_2012,blume-kohout_2013b,stark_2014,stephens_2021a}.

Recently it has been shown that there exist measurement techniques related to self-consistent tomography that are capable of detecting certain types of SPAM errors, without needing to explicitly reconstruct the state or measurement operators. Loop (or holonomic) SPAM tomography has been described theoretically for N-qubit systems \cite{jackson_2015,jackson_2017,jackson_2017b}, and demonstrated experimentally for 1- and 2-qubit systems \cite{mccormick_2017, feldman_2018}. It is capable of detecting correlated errors between state preparations and measurements, or between measurements performed in different locations. Loop SPAM tomography is powerful because the only assumption made about the states and the measurements is that their underlying Hilbert-space dimensions are known. It works by looking for self-consistency (or a lack thereof in the case of errors) in an over-complete set of measurements. 

Consider a quantum key distribution (QKD) system in which Alice and Bob wish to share a secret key. The expectation values of the measurements performed with their system can be described by
\begin{equation}\label{eq:trace}
	S^{ij} = \text{Tr}\left[\left(  {\hat a^i} \otimes {\hat b^j} \right) {\hat \xi} \right] .
\end{equation}
One possible protocol for such a QKD system is one that is based on Ekert's protocol, or something similar \cite{ekert_1991,brunner_2014}. In this case ${\hat \xi}$ describes the entangled state of a pair of particles that travel to Alice and Bob, ${\hat a^i}$ refers to Alice's $i$'th measurement operator, and ${\hat b^j}$ refers to Bob's $j$'th measurement operator. In this protocol there is a joint state, but the measurements are independent. Correlated errors can lead Alice and Bob to $think$ their generated key is secure, when in fact it is not. However, it has been demonstrated that loop SPAM tomography can detect these correlated errors, demonstrating to Alice and Bob the vulnerability of their system \cite{feldman_2018}. 

A different QKD protocol is one in which Alice and Bob prepare independent states, and then a joint measurement is performed on them. Equation~(\ref{eq:trace}) still applies, but in this case ${\hat \xi}$ describes the joint measurement, ${\hat a^i}$ refers to Alice's $i$'th state preparation, and ${\hat b^j}$ refers to Bob's $j$'th state preparation. This form of QKD is known as measurement-device-independent (MDI) QKD \cite{lo_2012}. The joint measurement can be performed by a third party, Charlie. It is assumed that Alice and Bob know their own states, but it can be shown that they do not need to know ${\hat \xi}$ (they do not need to trust Charlie) in order for their key to be secure. Recently, Moore and van Enk have proposed a method, related to self-consistent tomography, to determine if there are  correlated errors present in an MDIQKD system \cite{moore_2021}. Such errors could be due to an eavesdropper, or they could be due to more mundane, but hard to detect, hardware or software errors. Here we describe an experimental implementation of their method, using a form of MDIQKD known as twin-field (TF) QKD.

TFQKD was introduced as a way of extending the range of QKD systems \cite{lucamarini_2018,curty_2019}. Traditional QKD schemes have a maximum key rate that scales as the channel transmittance $\eta$, whereas the key rate for TFQKD scales as $\sqrt{\eta}$. There have been several experimental demonstrations of TFQKD, which have demonstrated its ability to transmit keys over ever larger distances, and in some cases have exceeded the repeaterless secret key capacity (the Pirandola-Laurenza-Ottaviani-Banchi, PLOB, bound)  \cite{minder_2019,wang_2019a,liu_2019a,zhong_2019,pittaluga_2021,wang_2022,chen_2022,pirandola_2017}. Here we implement a TFQKD system based on that of Zhong \emph{et al.} \cite{zhong_2019}. Our goal is not to implement a long-distance TFQKD system, but rather to show that we can use the technique proposed by Moore and van Enk to detect correlated errors in this system.

%%%%%%%%%%%%%%%%%%%%%%%%%%  Theory  %%%%%%%%%%%%%%%%%%%%%%%%%%
\section{Theory}\label{sec:thry}

\subsection{Correlated error detection}\label{sec:error}

Here we provide an overview of the theory behind our error detection technique; further details can be found in Ref.~\cite{moore_2021}. 
The operators describing the states produced by Alice and Bob can be written in the Pauli basis as
\begin{equation}\label{eq:ab}
	{\hat a^i} = \sum_{k} a_k^i { \hat \sigma_k}\:,\:\:\: {\hat b^j} = \sum_{\ell} b_{\ell}^j {\hat \sigma_{\ell}} ,
\end{equation}
where the $a_k^i$'s can be considered to be elements of a 4-dimensional vector $a^i$ that describes Alice's $i$'th state preparation, and the $b_{\ell}^j$'s similarly describe Bob's $j$'th state. The operator that describes the joint measurement $\hat \xi$ can be written as
\begin{equation}\label{eq:xi}
	{\hat \xi} = \frac{1}{4} \sum_{m,n} x_{mn}\: ({ \hat \sigma_m} \otimes { \hat \sigma_n}),
\end{equation}
where the $x_{mn}$'s are the elements of a $4\times4$ matrix $X$.
Substituting Eqs.~(\ref{eq:ab}) and (\ref{eq:xi}) into Eq.~(\ref{eq:trace}) and taking the trace yields 
% \begin{align}
% 	S_{ij} &= \text{Tr}\left[ \left( \sum_k \alpha_k^i \sigma_k \otimes \sum_{\ell} \beta_{\ell}^j \sigma_{\ell} \right) \sum_{m,n} x_{mn}\sigma_m\otimes\sigma_n \right] \\
% 	&= \text{Tr}\left[ \sum_{k,\ell,m,n} \alpha_k^i\beta_{\ell}^j x_{mn} \left(\sigma_k \otimes \sigma_{\ell} \right) \left(\sigma_m\otimes\sigma_n\right)\right] \\
% 	&=\sum_{k,\ell,m,n} \alpha_k^i \beta_{\ell}^j x_{mn} \text{Tr}\left[ \sigma_k\sigma_m \otimes \sigma_{\ell}\sigma_n \right].
% \end{align}
% Since the trace of the outer products is equal to the product of the traces, and for the Pauli matrices, $\text{Tr}\left[\sigma_i\sigma_j\right] = 2\delta_{ij}$, we can use
% \begin{equation}
% 	\text{Tr}\left[\sigma_i\sigma_j\otimes\sigma_k\sigma_{\ell}\right] = 4\delta_{ij}\delta_{k\ell}
% \end{equation}
% to get the result that
%
\begin{equation}\label{eq:Sij}
	S^{ij} =\sum_{k,\ell} x_{k\ell} a_k^i b_{\ell}^j.
\end{equation}
Here the expectation values $S^{ij}$ are the elements of a $4\times4$ matrix $S$. From Eq.~(\ref{eq:Sij}) we see that $S$ can be expressed as a matrix product:
\begin{equation}\label{eq:S}
	S = A^T{X}{B} ,
\end{equation}
where the 4 columns of ${A}$ and ${B}$ are the vectors that describe 4 different preparations of Alice's and Bob's states. If Alice ensures that the vectors that describe her 4 state preparations are linearly independent, we can multiply by the inverse
$\left(A^T \right)^{-1}$ to get
\begin{equation}\label{eq:inv}
	\left({A}^T\right)^{-1}{S} = {X}{B}.
\end{equation}

Suppose that Bob prepares 4 states that are described by the matrix $B$, and Alice prepares 4 states that are described by the matrix $A_1$. By Eq.~(\ref{eq:S}) they will measure the set of expectation values described by $S_1$. Now suppose that they perform a different set of measurements in which Bob keeps his state preparations the same, but Alice changes at least one of her state preparations. Alice's states are now described by the matrix $A_2$, and they measure expectation values $S_2$. Since $B$ is unchanged, the condition that there is a unique $X$ that is independent of Alice's state preparations is then, from Eq.~(\ref{eq:inv}),
\begin{equation}\label{eq:AS}
	\left({A_1}^T\right)^{-1}{S_1} = \left({A_2}^T\right)^{-1}{S_2}.
\end{equation}
We define the matrix $M$ as
\begin{equation}\label{eq:M}
	M \equiv \left({A_1}^T\right)^{-1}{S_1} - \left({A_2}^T\right)^{-1}{S_2}.
\end{equation}
From Eqs.~(\ref{eq:inv})-(\ref{eq:M}) it is apparent that the condition $M=0$ ensures that the measurements, described by $X$, are uncorrelated with Alice's state preparations. 

Alice can test whether or not there are any correlated errors between her state preparations and the measurements that Charlie is performing in the following way. Alice prepares five different states--four are used to construct the matrix $A_1$, and the fifth is combined with three of the first set to construct the matrix $A_2$. Bob prepares four different states, and Alice receives the measurement results for each trial from Charlie. From the measurements she can construct the matrices of expectation values $S_1$ and $S_2$. Note that Alice does not need to know what states Bob produces (she doesn't need to know $B$), but on each trial she does need to know which of the four states Bob is sending, so that she can properly arrange the measurements into $S_1$ and $S_2$. She does not need to trust the measurements that Charlie is performing (she doesn't need to know $X$). Alice constructs the matrix $M$ and if $M=0$, to within the uncertainty in the measurements, Alice can be confident that Charlie's measurements are independent of her state preparations. Furthermore, note that if Bob were also to prepare five different states, he could use the same process to test if Charlie's measurements are correlated with his state preparations.

Note that Alice may use as few as 5 state preparations. But if she and Bob decide on a QKD protocol that uses more states she can use as many as 8 preparations. The constraint is that each of the matrices $A_1$ and $A_2$ must contain independent columns so that they can be inverted.

\subsection{Twin-field quantum key distribution}\label{sec:tfqkd}

For our TFQKD system we use the encoding scheme described by Yin and Fu, which is essentially the same as that of Zhong \emph{et al.} with the addition of $Y$-basis states \cite{yin_2019,zhong_2019}. We use weak coherent states, and post-select all measurements on the presence of a single photon, so Alice and Bob's states consist of vacuum $\ket{0}$ and single-photon $\ket{1}$ contributions. The Bloch vectors that describe the states are distinguished by the phases of the corresponding coherent states. The $X$-basis states correspond to the phases $\theta=0,\pi$,
\begin{equation}\label{eq:X}
	\ket{+X} = \ket{\alpha} \:,\:\:\: \ket{-X} = \ket{\alpha e^{i\pi}} = \ket{-\alpha} ,
\end{equation}
while the $Y$-basis states correspond to the phases $\theta=\pi/2,3\pi/2$,
\begin{equation}\label{eq:Y}
	\ket{+Y} = \ket{\alpha e^{i\pi/2}} \:,\:\:\: \ket{-Y} = \ket{\alpha e^{i3\pi/2}} .
\end{equation}
The $Z$-basis states in this encoding scheme correspond to the vacuum and a phase-randomized weak coherent state. We do not need $\ket{+Z}=\ket{0}$ for our purposes. To keep the phase randomization practical, for $\ket{-Z}$ the phase is chosen from one of the four values $\theta = n \pi/2 \: (n = 0,1,2,3)$. Note that due to limitations of our hardware, we do not randomly choose the phase on every state preparation. Rather, we step through the phases in sequence. After averaging, the measured expectation value will be the same as if the phase was chosen randomly. While this would not work for a truly secure QKD system, it is sufficient for our purpose of demonstrating an error detection technique.

%%%%%%%%%%%%%%%%%%%%%%%%%%  Experiment  %%%%%%%%%%%%%%%%%%%%%%%%%%
\section{Experiments}

\subsection{Experimental design}

\begin{figure}[ht!]
\centering\includegraphics[width=7cm]{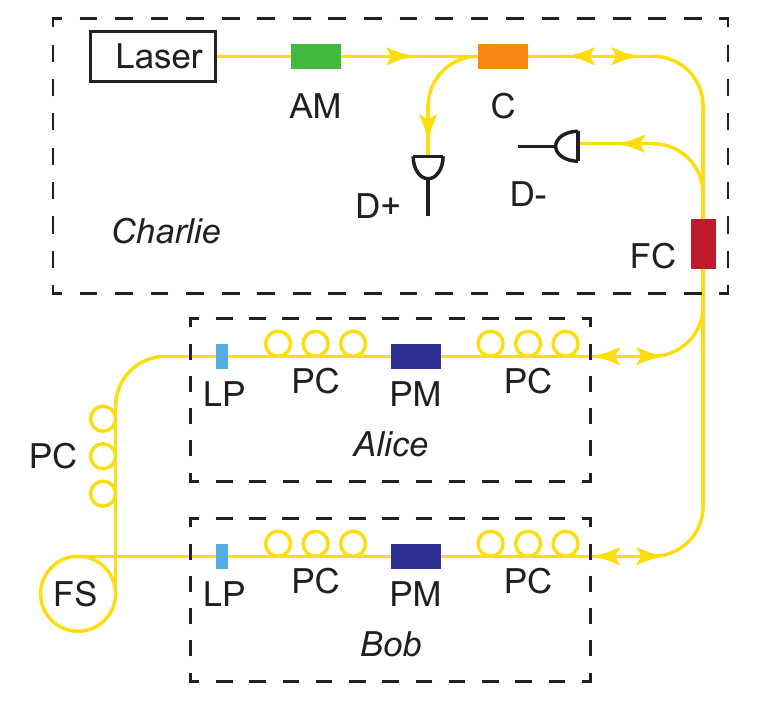}
\caption{\label{fig:app}The experimental apparatus. An amplitude modulator (AM) produces 1 ns pulses from a 1550 nm laser diode. These pulses pass through a circulator (C), and are split by a 2x2 fiber coupler (FC) before entering a Sagnac interferometer. Pulses travel in opposite directions around the Sagnac loop. Alice encodes information on the phase of the clockwise traveling pulse using a waveguide phase modulator (PM) synchronized with the arrival of her pulse, while Bob similarly encodes information in the phase of the counter-clockwise pulse. Polarization controllers (PC) and linear polarizers (LP) control the polarization in the interferometer; FS is a 40 m spool of optical fiber, which is used to introduce a time delay that ensures Alice's and Bob's state preparations are independent of each other. The pulses interfere at the coupler, and depending on their relative phases are detected with single-photon-counting modules at D+ or D-. By knowing which detector fires, Alice and Bob can generate a quantum key, or they calculate expectation values in order to perform error detection measurements.  }
\end{figure}

Our experimental apparatus, based on that of Zhong \emph{et al.}, is shown in Fig.~\ref{fig:app} \cite{zhong_2019}. Charlie generates 1~ns pulses at a rate of 1~MHz that he sends to Alice and Bob through single-mode optical fibers. Alice and Bob use phase modulators to select which of the basis states they transmit back to Charlie. To maintain temporal overlap and phase stability between Alice and Bob's states, the geometry is chosen to be that of a Sagnac interferometer, in which Alice's and Bob's photons propagate in opposite directions around a loop. Charlie uses single-photon counting modules (SPCMs) to detect the photons that return to him (the detected signal level is kept to <~0.01 photons per pulse). To eliminate dark and background counts the photon arrival times are recorded with time-to-digital converters, and only photons that arrive within 2.4~ns of the expected time are counted. 

%%%% Fig. 1.

\begin{figure}[ht!]
\centering\includegraphics[width=7cm]{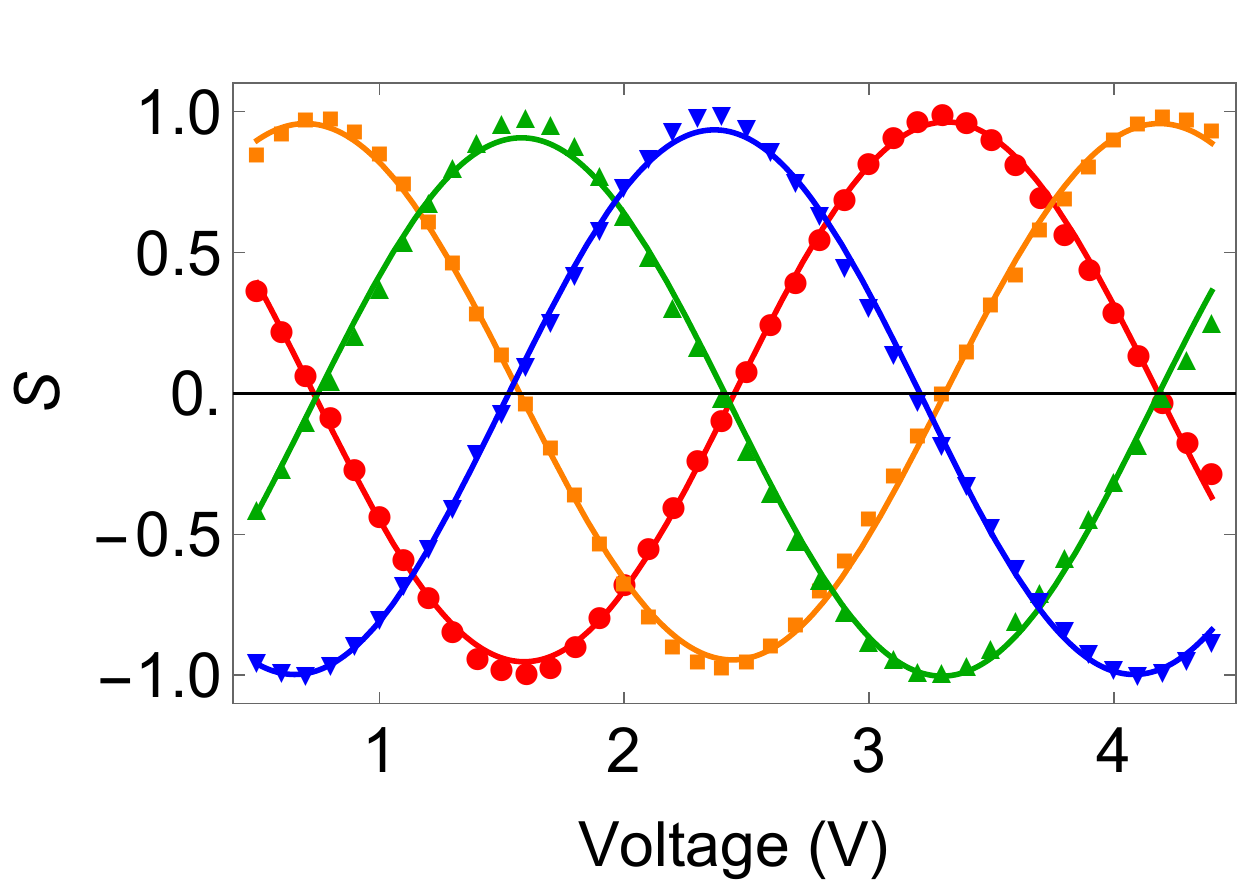}
\caption{\label{fig:pha}The expectation value $S$ as a function of the voltage applied Alice’s phase modulator. Markers
are measured, while the solid lines are sinusoidal fits to the data. Each curve corresponds to a
different phase shift applied by Bob: 0 (red~\textbullet), $\pi/2$  (orange~$\blacksquare$), $\pi$ 
({\textnormal {green}}~$\blacktriangle$), $3\pi/2$ (\textnormal{blue}~$\blacktriangledown$).}
\end{figure}

The expectation values for a given pair of Alice and Bob's state preparations are given by the number of detected photons $N_+$ and $N_-$ at detectors $D_+$ and $D_-$ as
\begin{equation}\label{eq:S2}
	S = \frac{N_+ - N_-}{N_+ + N_-}.
\end{equation}
Shown in Fig.~\ref{fig:pha} are plots of the expectation values as a function of the voltage that Alice applies to her phase modulator, for four different phase shifts applied by Bob. We see that the expectation values oscillate sinusoidally between $\pm 1$, as expected. Careful adjustment of the polarizations of the counter-propagating pulses is necessary to obtain the high-visibility interference that leads to this result. To calibrate the voltage that we need to apply to Alice's modulator to obtain a given phase shift, we fix Bob's phase shift, then we obtain 10 trials of $S$ as a function of voltage and fit each one to a sinusoidal function. We acquire these 10 trials during a time interval that is comparable to the time it takes for us to acquire the data we use to calculate $M$, in order to assure that any drift will be approximately the same for these different experiments. We then calculate the mean and standard deviation of the 10 sets of fit parameters. We find that the uncertainty in Alice's ability to produce a particular phase shift is $\pm~.029 \: \text{rad} \cong \pm~\pi/100$. This uncertainty determines the uncertainties in the elements of the matrix $A$ that describes Alice's state preparations.

%%%% Fig. 2

\subsection{Results}

On each shot Charlie reports to Alice and Bob which of his detectors have fired, and Bob reports to Alice a label that refers to which state he has produced. From this information, Alice can calculate expectation values of the measurements, and arrange them into the $S$ matrices as described above. Recall that Alice does not need to know Bob's states, but she does need enough information from Bob to distinguish between different states so that she can can properly construct the $S$ matrices. For each pair of Alice and Bob's state preparations Alice collects data from approximately $10^6$ shots. Alice also needs to know the uncertainty in $S$, so she repeats the above process 10 times, and calculates the mean and standard deviation of the elements of $S$. By knowing which states she produces on each shot, Alice can construct the $A$ matrices, including the uncertainty due to her ability to control the phase,  and finally she constructs the matrix $M$ and its uncertainty $\Delta M$. 
%If the ratio of the absolute value of the mean to the standard deviation $|M|/\Delta M$ is less than 1 for every matrix element, the mean is less than the uncertainty, so the uncertainty encompasses 0. Alice can then conclude that $M$ is 0 to within the uncertainty of the measurements, and there are no correlated errors. If this is not the case, then there are correlated errors present. (Note that we use $|M|/\Delta M$ to denote division that is performed matrix element by matrix element, not true matrix division.)

%%%%% Fig. 3
\begin{figure}[ht!]
\centering\includegraphics[width=7cm]{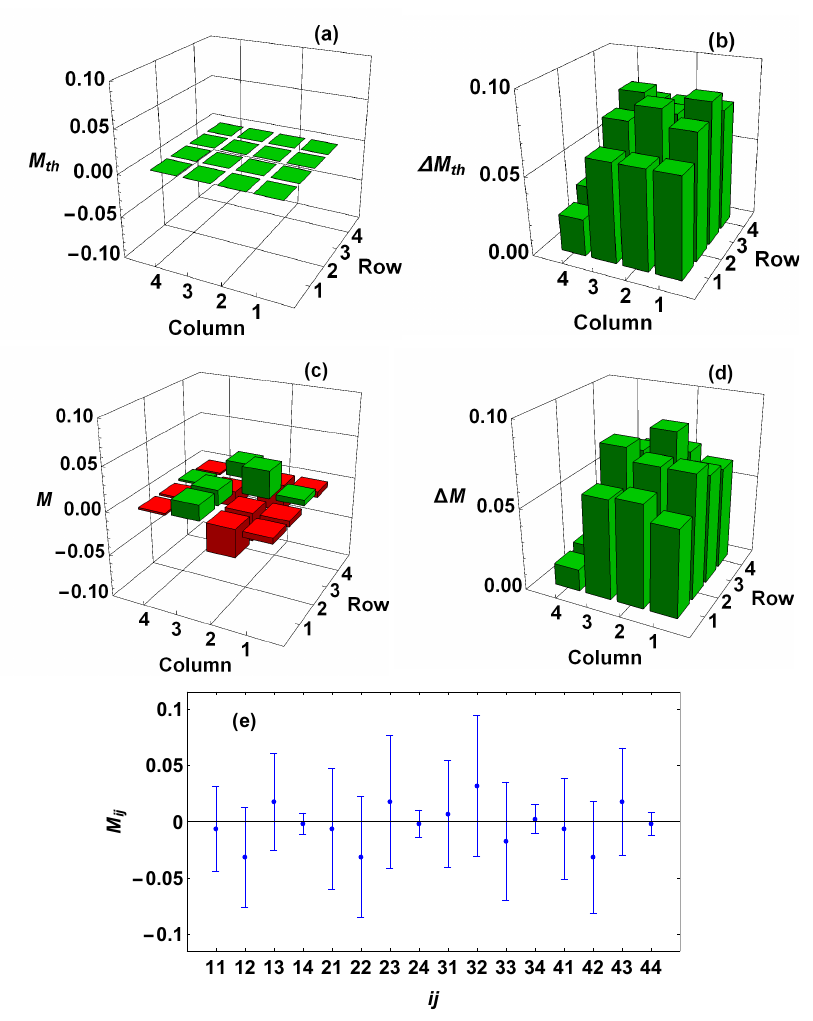}
\caption{\label{fig:noerr} Results corresponding to measurements when no errors have been introduced. The theoretically expected values for the mean $M_{th}$ and its standard deviation $\Delta M_{th}$ are shown in (a) and (b), while the corresponding experimentally measured values are sown in (c) and (d). A plot of the individual matrix elements $M_{ij}$ for the experimentally measured data is given in (e); here the error bars represent the 95\% confidence interval [Eq.~\eqref{eq:CI}]. It can be seen that the confidence interval for each matrix element contains 0. }
\end{figure}

Shown in Fig.~\ref{fig:noerr} are both theoretical and experimental matrices when we expect there to be no correlated errors present. The theoretical matrix $M_{th}$  is 0 for no errors [Fig.~\ref{fig:noerr}(a)], while the corresponding experimental matrix $M$ is nonzero due the the experimental uncertainties [Fig.~\ref{fig:noerr}(c)]. The theoretically expected uncertainties $\Delta M_{th}$ [Fig.~\ref{fig:noerr}(b)] are calculated by using the uncertainty in Alice's phase shift, and by assuming that the photocount statistics are Poissonian with a mean number of photons $N_{pho}$ equal to the experimental mean, and an uncertainty of $\sqrt{N_{pho}}$. We can see that the theoretically expected uncertainties are comparable to the experimental uncertainties [Fig.~\ref{fig:noerr}(d)].
%As seen in Fig.~\ref{fig:noerr}(c), the mean of every matrix element is less than the standard deviation, so all of the matrix elements are 0 to within the uncertainty of the measurements. As such no correlated errors are detected, which is what we would expect.

In order to determine whether or not the data shown in Fig.~\ref{fig:noerr} are consistent with $M=0$, we compute 95\% confidence intervals ($CI$) for all the elements in the experimentally determined $M$. These are given by
\begin{equation}\label{eq:CI}
	CI_{ij} = t^{*}_{N-1}\;\frac{M_{ij}}{\Delta M_{ij} / \sqrt{N}} .
\end{equation}
Here $N=10$ is the number of trials and $t^{*}_{N-1} = 2.262$ is the cutoff corresponding to the 95\% confidence interval of a $t$-distribution with $N-1$ degrees of freedom. As seen in Fig.~\ref{fig:noerr}(e), 0 is contained within the confidence interval for every matrix element. As such, our measurements are consistent with $M=0$, and hence no correlated errors have been detected.
%%%%% Fig. 4
\begin{figure}[ht!]
\centering\includegraphics[width=7cm]{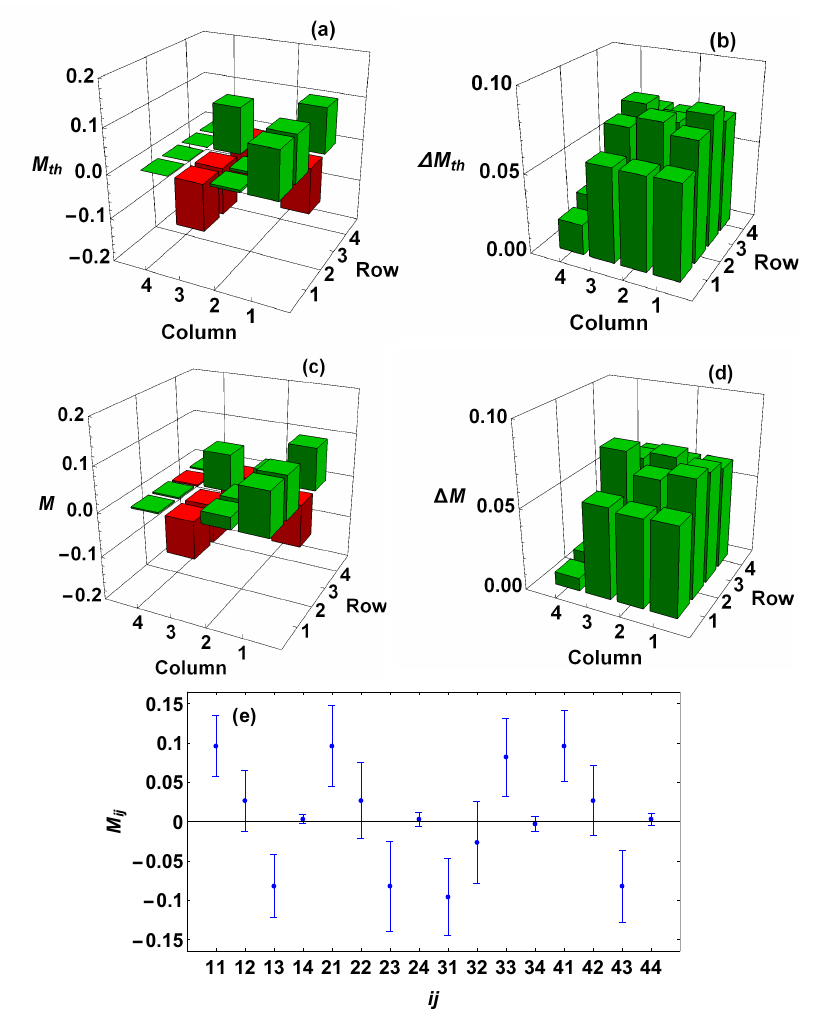}
\caption{\label{fig:err}Results corresponding to phase errors of $\pi/30$ introduced whenever Alice sets her state to $\ket{-X}$. The theoretically expected values for the mean $M_{th}$ and its standard deviation $\Delta M_{th}$ are shown in (a) and (b), while the corresponding experimentally measured values are sown in (c) and (d). A plot of the individual matrix elements $M_{ij}$ for the experimentally measured data is given in (e); here the error bars represent the 95\% confidence interval [Eq.~\eqref{eq:CI}]. It can be seen that the confidence intervals for several matrix elements do not include 0.}
\end{figure}

Shown in Fig.~\ref{fig:err} are results in which errors corresponding to a phase shift of $\pi/30$ on Alice's state preparations are introduced every time Alice prepares state $\ket{-X}$. The theoretical and experimental matrices and uncertainties are seen to be in good agreement. Figure~\ref{fig:err}(e) shows the individual matrix elements and their corresponding 95\% confidence intervals for the experimental measurements. Note that 0 is not contained within the confidence interval for several of the matrix elements.  The probability that we would find a particular value for $M_{ij}$ that is as at least as far from 0 as the measured value is given by the $p$-value of that matrix element, and we compute the $p$-value corresponding to each matrix-element. Even a single non-zero matrix element would mean $M\ne 0$. As such, we use the minimum $p$-value of the matrix elements (corresponding to the most extreme matrix element) to represent the $p$-value of the matrix.  For the data of Fig.~\ref{fig:err} we find the $p$-value to be $2 \times 10^{-8}$. %\textcolor{red}{[Albyn, this was calculated as follows. Find the 2-sided $p$-value of $t=\frac{M_{ij}}{\Delta M_{ij} / \sqrt{N}}$ .]}  
The probability that the data of Fig.~\ref{fig:err} is consistent with no error is quite low, and we have thus detected the $\pi/30$ errors on Alice's preparations of the state $\ket{-X}$ with high confidence.

In other experiments we have determined that we can also detect $\pi/30$ phase errors in the preparation of the states $\ket{+X}$ ($p=5 \times 10^{-8}$), $\ket{+Y}$ ($p=5 \times 10^{-5}$) and $\ket{-Y}$ ($p=7 \times 10^{-5}$) with similarly high confidence.  We have also found that we can detect larger phase errors in all of these states with even higher confidence. Note that we cannot detect phase errors in the state $\ket{-Z}$, as this corresponds to a state of random phase. Adding a constant phase shift to a random phase state has no effect on the state.

%%%%%%%%%%%%%%%%%%%%%%%%%%  Conclusions  %%%%%%%%%%%%%%%%%%%%%%%%%%
\section{Conclusions}

We have demonstrated that we can detect correlated SPAM errors in TFQKD using a technique related to self-consistent tomography. Specifically, Alice and Bob can share a secret key by preparing single-photon level coherent states, and using Charlie as an intermediary to report the results of joint measurements performed on their states. Alice can detect the presence of correlated errors without having to trust Charlie. All she needs is the information that Charlie makes publicly available, enough information from Bob to label his different state preparations (she does not need to know what the states actually are), and knowledge of her own state preparations.
We find that we can detect phase shift errors as small as $\pi/30$ in the state preparations with high confidence. It is reasonable that we should be able to detect phase errors of this size, given that our uncertainty in controlling the phase is $\pm \pi/100$.

While the experimental apparatus we have used was not intended to be a fully working TFQKD system, it does incorporate all of the major pieces of such a system. As such, we believe that our technique could be a useful tool for detecting correlated errors in TFQKD.

\begin{backmatter}
\bmsection{Funding}
National Science Foundation (NSF) (1855174).
% Content in the funding section will be generated entirely from details submitted to Prism. Authors may add placeholder text in the manuscript to assess length, but any text added to this section in the manuscript will be replaced during production and will display official funder names along with any grant numbers provided. If additional details about a funder are required, they may be added to the Acknowledgments, even if this duplicates information in the funding section. See the example below in Acknowledgements.

\bmsection{Acknowledgments}
We thank S. J. van Enk for helpful discussions, and for sharing a pre-publication version of Ref.~\cite{moore_2021}. We thank A. Jones for assistance with the statistical analysis of our data, L. Illing for the loan of some equipment, and J. Ewing for help with fabricating equipment. B. P. acknowledges support from the Gordon and Betty Moore Foundation.

\bmsection{Disclosures}
% Disclosures should be listed in a separate nonnumbered section at the end of the manuscript. List the Disclosures codes identified on the \href{https://opg.optica.org/submit/review/conflicts-interest-policy.cfm}{Conflict of Interest policy page}, as shown in the examples below:
%
% \medskip
%
% \noindent ABC: 123 Corporation (I,E,P), DEF: 456 Corporation (R,S). GHI: 789 Corporation (C).
%
% \medskip
%
% \noindent If there are no disclosures, then list ``
The authors declare no conflicts of interest.

% \bmsection{Data Availability Statement}
% A Data Availability Statement (DAS) will be required for all submissions beginning 1 March 2021. The DAS should be an unnumbered separate section titled ``Data Availability'' that
% immediately follows the Disclosures section. See the \href{https://www.osapublishing.org/submit/review/data-availability-policy.cfm}{Data Availability Statement policy page} for more information.

% OSA has identified four common (sometimes overlapping) situations that authors should use as guidance. These are provided as minimal models, and authors should feel free to
% include any additional details that may be relevant.

% \begin{enumerate}
% \item When datasets are included as integral supplementary material in the paper, they must be declared (e.g., as "Dataset 1" following our current supplementary materials policy) and cited in the DAS, and should appear in the references.

% \bmsection{Data availability} Data underlying the results presented in this paper are available in Dataset 1, Ref. [3].

% \bigskip

% \item When datasets are cited but not submitted as integral supplementary material, they must be cited in the DAS and should appear in the references.

% \bmsection{Data availability} Data underlying the results presented in this paper are available in Ref. [3].

% \bigskip

% \item If the data generated or analyzed as part of the research are not publicly available, that should be stated. Authors are encouraged to explain why (e.g.~the data may be restricted for privacy reasons), and how the data might be obtained or accessed in the future.

\bmsection{Data availability} Data underlying the results presented in this paper are not publicly available at this time, but may be obtained from the authors upon reasonable request.
%
% \bigskip

% \item If no data were generated or analyzed in the presented research, that should be stated.

% \bmsection{Data availability} No data were generated or analyzed in the presented research.
% \end{enumerate}

% \bmsection{Supplemental document}
% See Supplement 1 for supporting content. 

\end{backmatter}

%%%%%%%%%%%%%%%%%%%%%%% References %%%%%%%%%%%%%%%%%%%%%%%%%

%%%%%%%%%% If using BibTeX:
%\bibliography{references}

%%%%%%%%%% If preparing manually:
% \newcommand{\enquote}[1]{``#1''}

% \bibitem{Zhang:14}
% Y.~Zhang, S.~Qiao, L.~Sun, Q.~W. Shi, W.~Huang, L.~Li, and Z.~Yang,
%   \enquote{Photoinduced active terahertz metamaterials with nanostructured
%   vanadium dioxide film deposited by sol-gel method,}
%   {\protect\JournalTitle{Optics Express}} \textbf{22}, 11070--11078 (2014).

% {Optical Society}, \enquote{{OSA Publishing},}
%   \url{http://www.osapublishing.org}.

% \bibitem{FORSTER2007}
% P.~Forster, V.~Ramaswamy, P.~Artaxo, T.~Bernsten, R.~Betts, D.~Fahey,
%   J.~Haywood, J.~Lean, D.~Lowe, G.~Myhre, J.~Nganga, R.~Prinn, G.~Raga,
%   M.~Schulz, and R.~V. Dorland, \enquote{Changes in atmospheric consituents and
%   in radiative forcing,} in \enquote{Climate Change 2007: The Physical Science
%   Basis. Contribution of Working Group 1 to the Fourth assesment report of
%   Intergovernmental Panel on Climate Change,}  S.~Solomon, D.~Qin, M.~Manning,
%   Z.~Chen, M.~Marquis, K.~B. Averyt, M.~Tignor, and H.~L. Miler, eds.
%   (Cambridge University Press, 2007).

% \end{thebibliography}

\end{document}